# COMPLEXITY ANALYSIS OF NEXT-GENERATION VVC ENCODING AND DECODING

*Farhad Pakdaman[1], Mohammad Ali Adelimanesh[1], Moncef Gabbouj[2], Mahmoud Reza Hashemi[1]*

1 School of Electrical and Computer Engineering, University of Tehran, Tehran, Iran
2 Department of Computing Sciences, Tampere University, Tampere, Finland

**ABSTRACT**

While the next generation video compression standard, Versatile Video Coding (VVC), provides a superior compression efficiency, its computational complexity dramatically increases. This paper thoroughly analyzes this complexity for both encoder and decoder of VVC Test Model 6, by quantifying the complexity break-down for each coding tool and measuring the complexity and memory requirements for VVC encoding/decoding. These extensive analyses are performed for six video sequences of 720p, 1080p, and 2160p, under Low-Delay (LD), Random-Access (RA), and All-Intra (AI) conditions (a total of 320 encoding/decoding). Results indicate that the VVC encoder and decoder are 5x and 1.5x more complex compared to HEVC in LD, and 31x and 1.8x in AI, respectively. Detailed analysis of coding tools reveals that in LD on average, motion estimation tools with 53%, transformation and quantization with 22%, and entropy coding with 7% dominate the encoding complexity. In decoding, loop filters with 30%, motion compensation with 20%, and entropy decoding with 16%, are the most complex modules. Moreover, the required memory bandwidth for VVC encoding/decoding are measured through memory profiling, which are 30x and 3x of HEVC. The reported results and insights are a guide for future research and implementations of energy-efficient VVC encoder/decoder.

*Index Terms*— video coding, video decoding, complexity analysis, Versatile Video Coding (VVC), VVC Test Model (VTM)

## 1. INTRODUCTION

The demand for higher resolution video, diversity of devices available for record and playback of video, and multiple dimensions of video that increase the data rate, motivated the need for a new video coding standard with higher efficiency than the current High Efficiency Video Coding (HEVC) [1]. To respond to this urge, the Joint Video Experts Team (JVET), is developing the next generation video coding standard, Versatile Video Coding (VVC) [2], which is undergoing the final steps of standardization and is expected to be finalized soon in early 2020.

To gain better coding efficiency, several new coding tools and schemes have been integrated in VVC [2][3]. The Coding Tree Unit (CTU) with maximum size of 128×128 pixels and quadtree with a nested multi-type tree is one of the main improvements in this standard. Among other notable improvements are multiple core transforms, intra prediction with 65 directions, cross component linear model prediction, sub-block level motion vector prediction, affine motion compensated prediction, and new in-loop filters. Thanks to these new coding tools, VVC can already achieve more than 40% bitrate saving compared to HEVC.

While the main focus of the standardization committee is on coding efficiency, using new coding tools increases the complexity of both encoding and decoding. Therefore, employing such standard in consumer devices requires a thorough complexity analysis and needs assessment.

While the research for fast and power-efficient VVC encoding and decoding has already started [4][5][6], to the best of our knowledge, there are still no thorough complexity analysis of VVC encoder/decoder in the literature. Thus, this paper presents a thorough complexity analysis of VVC encoding and decoding operations, to quantify different aspects of their complexity. While the VVC Test Model (VTM) [3] is not optimized for complexity and cannot perform real time encoding/decoding, test models are often accepted implementations for complexity analysis, and performance assessment of low complexity algorithms. For the HEVC standard, HEVC Test Model (HM) [7] has been used in [8] and [9] for complexity analysis of video encoding and decoding, respectively. For VVC, the current assessments only concentrate on coding efficiency [10], subjective quality [11], encoding of intra-frames [12], or the energy of decoding [13].

To provide an extensive complexity analysis of VVC, this paper analyzes both encoders and decoders of VVC and HEVC. These analyses quantify the share of each coding/decoding module, and measures the complexity of newly introduced coding tools in VVC. Moreover, the total computational complexity and memory bandwidth requirements for VVC encoding and decoding are measured and compared to those of HEVC. Discussions are provided that detail how the complexity changes with change of QP, configuration, and resolution. The analyses presented in this paper can be a guide for future research and development of low complexity and energy-efficient VVC encoding/decoding.

To do so, this paper considers the three coding configurations suggested by [14] for VVC and by [15] for HEVC, i.e. Low delay (LD), Random Access (RA) and All-Intra (AI) configurations. Six different video sequences with resolutions of 1280×720, 1920×1080, and 3840×2160 pixels were encoded with VTM 6 and HM 16, with Quantizer Parameters (QP) of 22, 27, 32, and 37, and the three coding configurations. Next, all the coded videos were decoded with VTM or HM decoders. Intel VTune Amplifier 2019

[16] was used to analyze the complexity of a total of 320 coding or decoding operations, on an Intel core i7 4790K machine with maximum clock speed of 4 GHz, 8 GBs of Memory, and Windows 10 platform. These analyses measure the complexity of each coding module of VVC, and compare them with corresponding modules of HEVC. Moreover, memory profiling was performed to assess the required memory bandwidth for VVC encoding and decoding, and to compare it with HEVC. The detailed VTune outputs, all encoded videos, and detailed tables including complexity break-down for each video sequence, are publicly available in [17] and also [18].

The rest of the paper is organized as follows. Section 2 summarizes some of the new coding tools of VVC, sections 3 and 4 present the detailed analysis of VVC encoding and decoding, respectively. Finally, section 5 provides conclusion and discussions.

## 2. VERSATILE VIDEO CODING

To gain more compression efficiency, VVC introduces several improvements and new coding tools compared to previous standards. This section summarizes some important modules.

**Block Partitioning:** VVC extends the concept of CTU in HEVC. A CTU can be as large as 128×128 pixels and is partitioned with a quadtree with nested multi-type tree (QTMT) scheme. This allows splitting a block into square, binary, or horizontal and vertical ternary sub-units. This structure unifies the concepts of Coding Unit (CU), Prediction Unit (PU), and Transform Unit (TU) into CU. Such flexibility allows a detailed modeling of video content.

**Intra Prediction:** VVC uses 67 intra modes, consisting of 65 directional modes, DC, and planar. Intra prediction is performed for CUs of up to 64×64 pixels, and unlike HEVC, non-square blocks are available in intra prediction as well. While intra angular modes support directions from 45° to -135°, to support non-square blocks an adaptive scheme is used to use wide-angle modes. Moreover, an intra smoothing filter is used that adaptively chooses a four-tap filter based on the directional mode. To further reduce the cross-component redundancy, VVC employs a linear prediction mode scheme that allows chroma samples to be predicted based on reconstructed luma samples. Another interesting improvement is that VVC extends the reference samples by enabling the use of several reference lines, which improves the quality of prediction.

**Inter prediction:** The motion in VVC can be signaled via explicit transmission of motion parameters, via skip mode, or via merge mode which includes deriving motion parameters from spatial and temporal candidates. The merge mode uses candidates from spatial, temporal, and zero Motion Vector (MV) like HEVC, plus a pairwise average vector, and a history-based vector from a FIFO table. A new scheme named merge mode with motion vector difference (MMVD) is introduced that refines the derived motion via a motion vector difference (MVD). Moreover, symmetric MVD coding is introduced, where the reference picture indices of list-0 and list-1, and the MVD of list-1 are derived at the decoder.

To compensate different kinds of non-translational motion in both merge mode and affine Advanced Motion Vector Prediction (AMVP), VVC employs an affine motion compensation prediction. The motion is signaled via motion information of two or three control points from the corners of the block. At the decoder side, the motion for each location is inferred based on these information. While HEVC uses a constant quarter-luma-sample accuracy for signaling MVD, VVC uses an adaptive MV resolution that allows a MV resolution between four to 1/16 luma-samples depending on the mode. Triangle partitioning, decoder side motion vector refinement, and bi-directional optical flow are among other mentionable coding tools.

**Table 1** Details of video sequences used for complexity analysis

| Sequence | Resolution | Frame rate | Bit depth | Tested configs |
|---|---|---|---|---|
| BQTerrace | 1920×1080 | 60 | 8 | AI, LD, RA |
| BasketballDrive | 1920×1080 | 50 | 8 | AI, LD, RA |
| Cactus | 1920×1080 | 50 | 8 | AI, LD, RA |
| Johnny | 1280×720 | 60 | 8 | AI, LD, RA |
| KristenAndSara | 1280×720 | 60 | 8 | AI, LD, RA |
| DaylightRoad2 | 3840×2160 | 60 | 10 | AI, LD |

**Transform:** VVC uses transforms of up to 64×64 for luma and 32×32 for chroma samples which suits higher resolution samples. For the maximum transform size, a high-frequency zeroing scheme retains only the lower-frequency half of coefficients. A multiple transform selection scheme allows selecting the best horizontal and vertical transform cores, among different DCT and DST cores. To further exploit the spatial redundancy, a secondary transform is introduced that uses 4×4 and 8×8 non-separable transforms. A new dependent scalar quantization is used, in which the set of admissible reconstruction values depends on previously reconstructed coefficients.

**Entropy coding:** VVC uses an improved Context-Adaptive Binary Arithmetic Engine (CABAC). Unlike HEVC, coding of transform coefficient levels is dependent only upon TU size, which results in various options for selection of coefficient groups. A new QP-dependent context model initialization is introduced. Moreover, the selection of probability models for syntax elements depends on value and number of non-zero elements in a local neighborhood.

**In-Loop filters:** VVC uses three in-loop filters. While Deblocking Filter (DBF) and Sample Adaptive Offset (SAO) are very similar to those of HEVC, Adaptive Loop Filter (ALF) is only used in VVC. ALF uses a block classification scheme to choose from 25 different filters, based on the direction and level of local gradients. This filter is applied after DBF and SAO, on blocks of 4×4 pixels.

## 3. ENCODER COMPLEXIY ANALYSIS

This section analyzes the computational complexity and memory requirements of VVC encoding, measures complexity for each encoding tool, and compares them with HEVC. Findings show how complexity changes with resolution, configuration, and QP.

**Complexity break-down:** Fig. 1 presents the complexity break-down of VVC encoder, for the average of all test sequences used in this study (for the sake of conciseness, the average is presented here and the detailed break-down for each video sequence, as well as all VTune reports, and encoded videos, are available on [17] and [18]). These video sequences are introduced in Table 1. The total complexity is broken down into six categories of Motion Estimation (ME), Intra Prediction (IP), Transform and Quantization (T/Q), Entropy Coding (EC), Loop-Filters (LF), and Memory (Mem) operations. While the results have some variation based on different sequences, for LD and QP=22 each above category has 47%, 6%, 28%, 10%, 2%, and 3% of the total complexity, respectively. For QP=37 this break-down is 57%, 4%, 17%, 5%, 9%, and 3%. In other words, in lower bitrates the shares of T/Q and EC decrease, and the shares of ME and especially LF increase. While RA has a similar trend to LD, for AI on average, IP with 29%, T/Q with 44%, and EC with 15% consume the majority of total complexity. Another observation is that for higher resolutions, the share of ME decreases and instead, the shares of

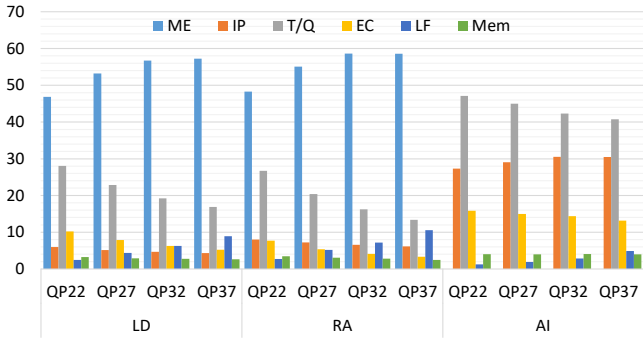
**Fig. 1** VVC encoder complexity break-down

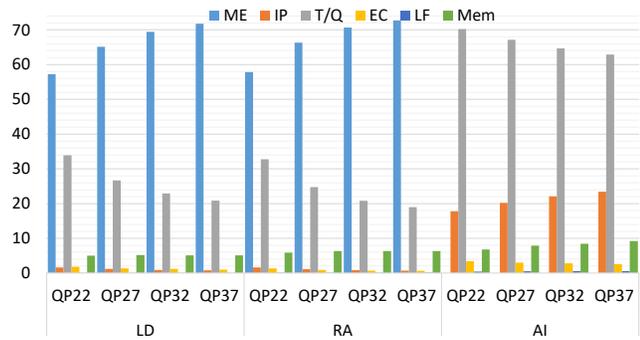
**Fig. 2** HEVC encoder complexity break-down

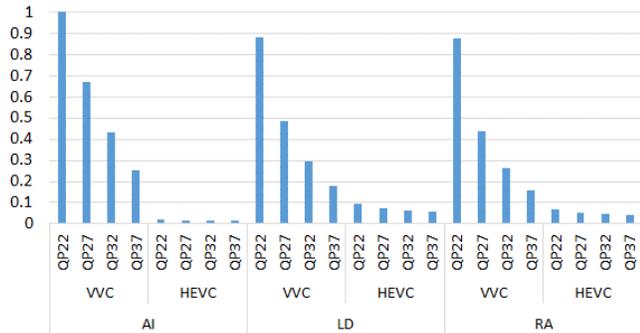
**Fig. 3** Normalized average complexity of VVC and HEVC Encoding

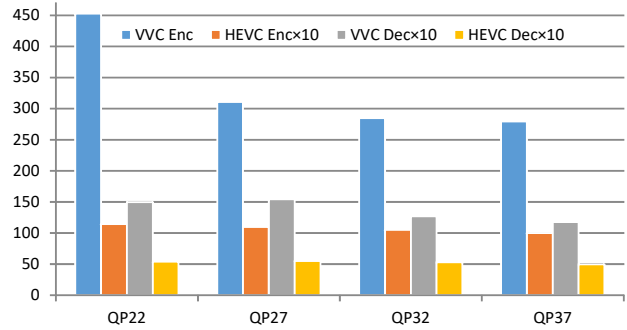
**Fig. 4** Comparison of average memory bandwidth requirements for VVC and HEVC (GB/s)

EC and T/Q increase.

It is observed that ME has the highest share in LD and RA. Among the sub-parts of ME, affine search with an average of 17%, affine AMVP with 3%, pattern search with 7%, fractional pattern with 7%, merge mode with 14%, triangle merge with 2%, and AMVP with 2% are the most important ones (for LD).

A major part of VVC encoding complexity is due to the recursive calling of coding functions through the QTMT structure. This complexity is distributed through all coding functions and cannot be considered as a separate coding tool. However, through limiting the coding structure one can measure its effect on the overall coding complexity. In an experiment, the LD coding was repeated for coding depths (quad-tree + multi-type tree) of 1 to 4 and compared to the default, which is 6. This reduced the overall complexity by 82%, 76%, 58%, and 26%, respectively. Moreover, turning the multi-type splitting off, has led to 83% time saving (of course both experiments had major effect on coding efficiency.)

**Comparison with HEVC:** Similarly, the average break-down for HEVC encoder is presented in Fig. 2. Comparing Fig. 1 and Fig. 2, it can be observed that LF consumes a larger share in VVC. For LD its share has been increased from an average of 0.1% in HEVC to 5.5% in VVC. Also, the share of IP in LD has increased from 1% in HEVC to 5% in VVC. While a similar trend is true for EC, ME and T/Q's share have decreased in VVC by a factor of 0.8x. As ME has actually become more complex and energy consuming in VVC, this observation shows that other coding modules (e.g. IP, EC, and LF) had more dramatic increases.

Moreover, Fig. 3 compares the total complexity of HEVC and VVC encoding based on normalized average of all 720p and 1080p videos. On average in LD, RA, and AI configurations, VVC takes 5x, 7x, and 31x the time of HEVC encoding. An interesting observation is that unlike previous standards where intra-coding was much simpler than inter-coding, intra-coding of VVC takes over the LD and RA by 1.3x and 1.4x, respectively. Moreover, it can be seen that the complexity of VVC encoding depends more on QP compared to HEVC. For LD, encoding with QP=22 is 4.8x the complexity of encoding with QP=37, while this value for HEVC is only 1.6x. This means that encoding with a higher quality requires more processing as well as higher bitrate.

**Memory bandwidth analysis:** To have a better perspective of VVC and HEVC encoding/decoding requirements, the required memory bandwidth for encoding and decoding have been measured for 1080p videos, and LD configuration. As neither of VTM and HM can perform real time encoding or decoding, their total accessed memory were measured and scaled to the real time encoding. Fig. 4 reports these values for both encoding and decoding (HEVC Enc/Dec, and VVC Dec are multiplied by 10 in Fig. 4 for better readability). Due to the higher amount of information to process, both encoding and decoding require more memory bandwidth in lower QPs. The average values for VVC encoding and decoding are 332 GB/s and 14 GB/s. These values for HEVC encoding/decoding are 11 and 5 GB/s respectively. This increase is due to several factors such as larger maximum CU size and more prediction candidates, which require more storage; and also, more diverse CU partitions and various coding modes, which lead into multiple and irregular accesses to the same data.

While VTM and HM are not optimized for computation and memory efficiency, this comparison shows the increased requirements of VVC implementation compared to HEVC. Both hardware architecture techniques [19][20] and algorithm level optimizations [21] proposed for previous standards can be extended to mitigate this issue in VVC. As an example, authors in [22] reduced the required bandwidth of 14.83 GB/s for a specific HEVC encoder chip, to 2.97 GB/s, by designing an efficient on-chip memory system and a data reuse scheme [23].

## 4. DECODER COMPLEXIY ANALYSIS

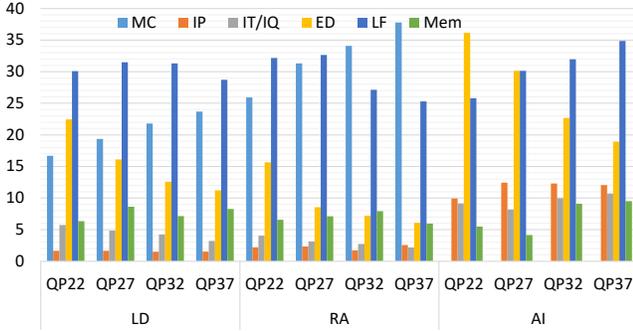

**Fig. 5** VVC decoder complexity break-down

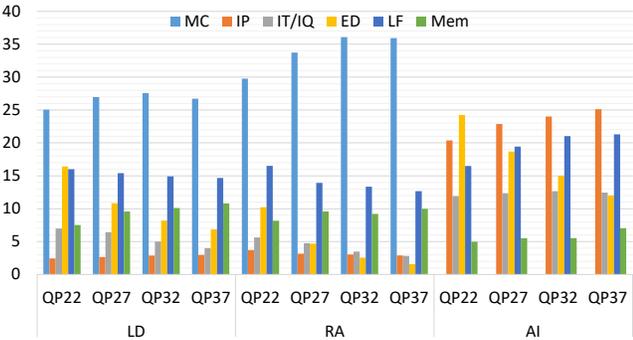

**Fig. 6** HEVC decoder complexity break-down

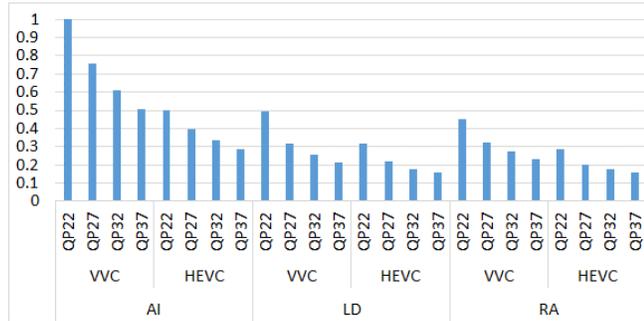

**Fig. 7** Normalized average complexity of VVC and HEVC decoding

This section analyzes the complexity of VVC decoding, measures complexity for each decoding tool, and compares with HEVC. Findings show how complexity changes with resolution, configuration, and QP.

**Complexity break-down:** Fig. 5 presents the average complexity break-down of VVC decoder (for the sake of conciseness, the average is presented here and the detailed break-down for each video sequence, as well as all VTune reports, and encoded videos, are available on [17] and [18]). Similar to VVC encoding, the results have some variations for different sequences. However, for LD and QP=22, LF, Motion Compensation (MC), Entropy Decoding (ED), Mem, Inverse Transform and Inverse Quantization (IT/IQ), and IP, with 30%, 17%, 22%, 6%, 6%, and 2% comprise most of the complexity. These values for QP=37 are 29%, 24%, 11%, 8%, 3%, and 2%, respectively. In other words, in larger QPs, the share of MC increases, and shares of ED and IT/IQ decrease. This is partly because lower QP videos have more bits to process. Another relevant observation is that for larger resolutions, the share of MC and LF decrease and ED increases. While RA is similar to LD, for AI on average, LF with 31%, ED with 27%, IP with 12%, IT/IQ with 10%, and Mem with 7% are the most demanding operation.

A major part of VVC decoding is spent on LFs which consist of SAO with 3%, ALF with 13%, and DBF with 17% of total complexity on average. Moreover, interpolation filters consume 10% of complexity and are a sub-part of MC.

**Comparison with HEVC:** Similar average break-down for HEVC decoding is presented in Fig. 6. Comparing with Fig. 5, it can be observed that ED and LF's complexity have increased from 5% and 15% in HEVC, to 10% and 32% in VVC (for LD). The share of MC in LD also decreases from 26% in HEVC to 20% in VVC. Moreover, in AI, IP's share has decreased from 23% in HEVC to 12% in VVC. Analogous to the case of encoding, as intra prediction and motion compensation have actually become more complex, these observations show that the increase of other coding tools have been more dramatic.

Fig. 7 compares total complexity of VVC decoding with HEVC decoding in different QPs and configurations, for normalized average of 1080p and 720p videos. VVC has 1.5x, 1.5x, and 1.8x the complexity of HEVC decoding in LD, RA, and AI, respectively. The total complexity of AI compared to LD and Ra has also increased. For VVC decoding in AI on average is 2.2x of LD and RA, while for HEVC this value is 1.7x and 1.8x respectively. These results again show that new techniques in intra-coding have made it much more complex. Moreover, decoding with lower QPs requires more processing in both standards. For LD, decoding with QP=22 is 2.3x times the complexity of QP=37. This value for HEVC is 2x times.

## 5. CONCLUSION AND DISCUSSION

An extensive complexity analysis of VVC encoding and decoding was presented in this paper. Share of different operations in encoding and decoding was reported for 6 video sequences and compared with those of HEVC. It was observed that VVC encoding and decoding are 5x and 1.5x of HEVC in LD, and 31x and 1.8x in AI. This suggests that fast intra prediction schemes should be explored for simplification of both intra direction estimation and fast CU partitioning. Due to the similarity of concepts, many successful proposals for HEVC can be extended for fast intra prediction in VVC [24][25][26].

It was observed that in LD and RA coding, new ME tools consume a major part of total complexity. These tools are the merge mode with extended candidates, affine ME, triangle merge mode, and pattern searches. While individual fast prediction modes are of interest to accelerate the process of individual modes [27][28], a smart mode decision scheme to prune unlikely modes based on the context can achieve major complexity reduction.

In decoding, MC, LF, ED, and IT/IQ consume the major portion of processing power. An important part of increased complexity in VVC decoding belongs to three LFs that follow reconstruction. Hence, optimized implementations [29][30] and parallel processing [31][32] can be used to decrease the complexity of VVC decoding.

In addition to the computational complexity, VVC requires a large memory bandwidth to access data. It was reported that VVC encoding and decoding use 30x and 3x memory bandwidth of the HEVC. As ME operations and CTU partitioning are accountable for most of these access, existing approaches such as starting point estimation for ME [21][33], software-managed memories [19], data reuse [23], and reference frame compression [20] should be explored to make VVC encoding/decoding affordable.